# Unambiguous determination of structure parameters for soft matter samples made possible with polarization analysis on JCNS SANS using a $^3$He spin filter

E. Babcock[1], A. Ioffe[1], A. Radulescu[1], V. Pipich[1]

[1]*Juelich Centre for Neutron Science at FRM 2, Forschungszentrum Juelich GmbH, 85747 Garching, Germany.*

Incoherent background can create an intrinsic problem for standard small angle neutron scattering measurements. Biological samples contain hydrogen which is a strong incoherent scatterer thus creating an intrinsic source of background that makes determination of the coherent scattering parameters difficult in special situations. This can especially be true for the Q-range from around 0.1-0.5 Å$^{-1}$ where improper knowledge of the background level can lead to ambiguity in determination of the samples structure parameters. Polarization analysis is a way of removing this ambiguity by allowing one to distinguish the coherent from incoherent scattering, even when the coherent scattering is only a small fraction of the total scattered intensity. $^3$He spin filters are ideal for accomplishing this task because they permit the analysis of large area and large divergence scattered neutron beams without adding to detector background or changing the prorogation of the scatter neutron beam. This rapid note describes the application of $^3$He neutron spin filters, polarized using the spin-exchange optical pumping method, for polarization analysis on a protein sample to unambiguously extract the coherent scattered intensity.

Using SANS to probe the structure of biological samples requires precise determination of the parameters used to model or fit the scattering pattern generated by the sample. These parameters include forward scattering, structure, size, and background. The first three parameters are related to coherent scattering and the fourth arises from incoherent scattering, the coherent scattering gives information on the samples structure whereas the incoherent scattering is just background competing with the desired information. Normal biological samples contain hydrogen which produces incoherent scattering, thus the sample itself creates an intrinsic background.

Fig. 1 shows a cartoon example of how this background leads to ambiguity in the parameters determined from a given protein sample. In the models or fits of such data, the parameter for the background is coupled to the other parameters. Changing the assumed level of background can infer very different form factors, i.e. the samples shape, which comes from the power law of the scattering, which then couples to the determination of other important parameters such as the samples size. Essentially any of the possible ranges for the form factor can be fit to the curve, which was generated for a Q$^{-4}$ dependence, i.e. Porod like scattering. The information on form factor is entirely determined in the fit by the data points in the Q>10$^{-1}$Å$^{-1}$ regime for this simulated data. For real data this regime would have low statistical certainty because of the low scattering cross section, and accurate information is hard to obtain. Accurate information in this Q-range is also very crucial for modeling protein form with Monte Carlo simulation.

Insufficient knowledge of the background can lead to ambiguity. Since coherent scattering does not change a scattered neutrons polarization, whereas incoherent scattering does, analyzing the scattered neutron spins of a polarized incident neutron beam allows one to distinguish between the coherent and incoherent components, eliminating the ambiguity from background. For SANS the polarization analysis has strict requirements of covering a highly divergent beam and not altering the neutron propagation direction or adding to detector background. $^3$He spin filters are ideally suited to this task [1].

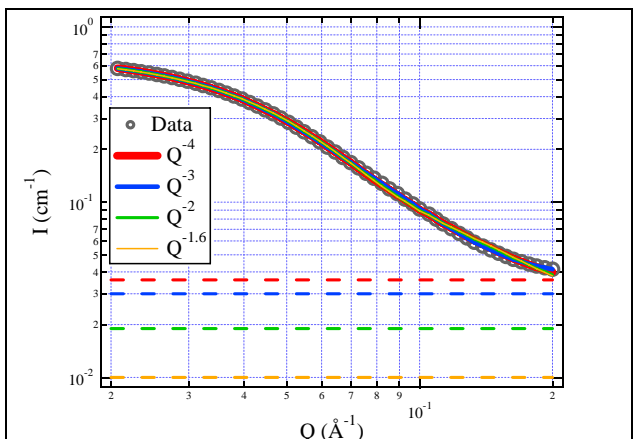

Fig. 1. A simulated scattering curve, open grey circles, with a $Q^{-4}$ form factor. The four solid lines, which are nearly impossible to distinguish because they overlap, red, blue green, and orange, are Beacauge fits with the form factor held to power laws of -4, -3, -2, and -1.6 respectively. The dotted lines are the corresponding value of the background from the individual fits of the same color. From data like this alone, accurate determination of the form factor would not be possible.

Incident beam polarization on KWS2 was available for experiments for several weeks at the end of the 23$^{rd}$ reactor cycle at the FRM II. The beam polarization was accomplished with a single super mirror in transmission geometry providing an incident beam polarization of 97% at 4.5Å, a Mezei-style spin flipper was used before the sample to flip the incident spin. To provide polarization analysis (PA), a polarized $^3$He neutron spin filter was placed after the sample. This $^3$He neutron spin filter gave a high polarization efficiency of 95% at the beginning of the experiments and was 93% after over 36 hours of measurements.

To achieve this high performance, the $^3$He cell and the sample were in the common magnetic environment of a uniform 10 G field provided by a 38 cm long shielded end-compensated solenoid. A second similar shielded solenoid was used for isolation from the external field gradients created by the spin flipper, and for adiabatic rotation of the vertical neutron guide field into the longitudinal direction of the solenoids. This arraignment provided very high performance for the $^3$He, giving a total relaxation time of 430 hours for the $^3$He polarization during the measurements. A picture of the installation is shown in figure 2.

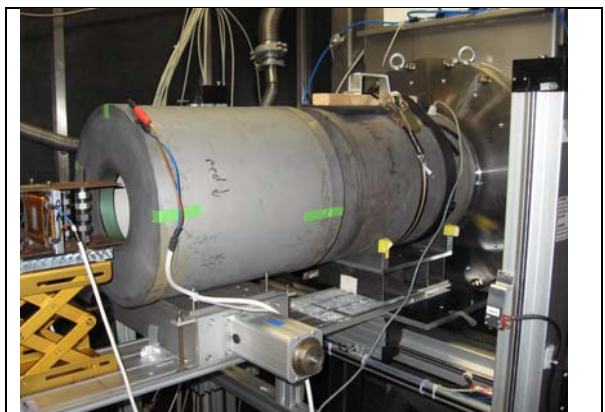

Fig. 2. Picture of the installation on KWS2 showing the two solenoids and the spin flipper. The sample is inside the solenoid on the right a few cm before the $^3$He spin filter cell. Overlap from the vertical field of the spin flipper and the longitudinal field of the first (left) solenoid allow adiabatic rotation of the neutron spin.

Polarization of the cell was carried out in the JCNS SEOP laboratory. The initial $^3$He polarization of 72% was the saturated maximum value attainable for the cell used. This level of performance would allow us to operate for 4 to 5 days on a single polarization of the cell with good neutron performance. Since the measurement times for each sample were found to be on the order of an hour or less, many experiments could be performed in this time.

A sample plot of the data obtained is shown in figure 3. Curves for both standard un-polarized measurements, and measurements using PA are shown. The data with polarization analysis clearly demonstrates the removal of the incoherent background and the data can be fit to a model that only contains the coherent scattering components. While this is a straightforward example, for a high-concentration sample (5% protein in deuterated buffer solution) with $Q^{-4}$ structure factor, we expect that the technique could become particularly useful in difficult situations. When the Q-range is insufficient to observe the level of the background, the structure factor is a lower order power, the sample cannot be measured at high concentrations due to aggregation, or when possible hydrogen exchange between the sample and solvent makes correct solvent subtraction and background determinations problematic, PA could provide the needed strait forward solution. Additionally, PA may also enable measurements with only one sample concentration. Other types of samples may require more complex models to describe the data where removal of the background will similarly remove ambiguity by allowing modeling containing only the coherent scattering.

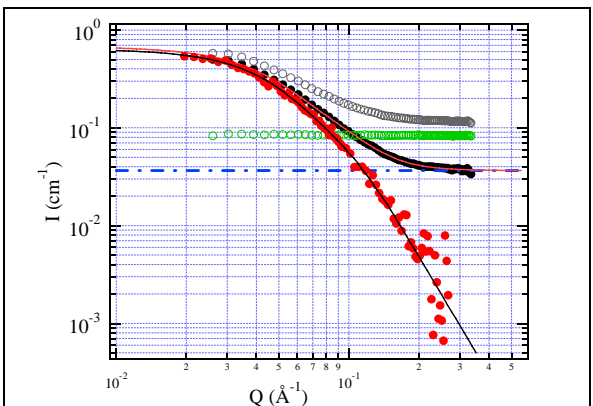

Fig. 3. Sample of the data obtained. The grey/green markers are the data obtained from standard measurements for the sample+solvent and a separate measurement of the solvent respectively. After data treatment one obtains the black/red markers which are the standard SANS signal, and the signal with PA respectively. The red/black lines are fits to the data, and the blue dotted line is the presumed background level from the fit of the standard data. Fitting the PA data to a Beaucage fit with no background, one can obtain the structure dependent fit parameters with no additional assumptions, and the values obtained are consistent with those obtained from standard measurements using several concentrations and no PA. Thus knowledge of the protein is obtained unambiguously with the one measurement employing PA. The data shown is from a sample prepared and by C. Sill who assisted the measurements.

We would like to thank T.R. Gentile and W.C. Chen for the loan of the cell (called Bullwinkle) used in the measurement. Future plans will focus on developing an optimal magnetic cavity for the J1 cell, made in FZ-Juelich [2], to allow in-situ polarization and increase the available Q-range to 0.5-0.6 $\text{Å}^{-1}$ while using standard sample environment. Further, installing a new super mirror polarizer will provide about a 100x higher flux on the sample, increasing statistical accuracy and lowering measurement time.